\documentclass[sigconf]{aamas} 

\usepackage{balance} 
\usepackage{algorithm}
\usepackage{algorithmic}
\usepackage{marvosym} 
\usepackage{hyperref}
\usepackage{enumitem} 

\setcopyright{ifaamas}
\acmConference[AAMAS '24]{Proc.\@ of the 23rd International Conference
on Autonomous Agents and Multiagent Systems (AAMAS 2024)}{May 6 -- 10, 2024}
{Auckland, New Zealand}{N.~Alechina, V.~Dignum, M.~Dastani, J.S.~Sichman (eds.)}
\copyrightyear{2024}
\acmYear{2024}
\acmDOI{}
\acmPrice{}
\acmISBN{}

\acmSubmissionID{<<EasyChair submission id>>}

\title[AAMAS-2024 Formatting Instructions]{Measuring Policy Distance for Multi-Agent Reinforcement Learning}

\author{Tianyi Hu}
\affiliation{
  \institution{Institute of Automation, CAS}
  \institution{School of Artificial Intelligence, UCAS}
  \city{Beijing}
  \country{China}}
\email{hutianyi2021@ia.ac.cn}

\author{Zhiqiang Pu}
\affiliation{
  \institution{Institute of Automation, CAS}
  \institution{School of Artificial Intelligence, UCAS}
  \city{Beijing}
  \country{China}}
\email{zhiqiang.pu@ia.ac.cn}

\author{Xiaolin Ai}
\affiliation{
  \institution{Institute of Automation, CAS}
  \city{Beijing}
  \country{China}}
\email{xiaolin.ai@ia.ac.cn}

\author{Tenghai Qiu}
\affiliation{
  \institution{Institute of Automation, CAS}
  \city{Beijing}
  \country{China}}
\email{tenghai.qiu@ia.ac.cn}

\author{Jianqiang Yi}
\affiliation{
  \institution{Institute of Automation, CAS}
  \institution{School of Artificial Intelligence, UCAS}
  \city{Beijing}
  \country{China}}
\email{jianqiang.yi@ia.ac.cn}

\begin{abstract}

Diversity plays a crucial role in improving the performance of multi-agent reinforcement learning (MARL). Currently, many diversity-based methods have been developed to overcome the drawbacks of excessive parameter sharing in traditional MARL. However, there remains a lack of a general metric to quantify policy differences among agents. Such a metric would not only facilitate the evaluation of the diversity evolution in multi-agent systems, but also provide guidance for the design of diversity-based MARL algorithms. In this paper, we propose the multi-agent policy distance (MAPD), a general tool for measuring policy differences in MARL. By learning the conditional representations of agents' decisions, MAPD can computes the policy distance between any pair of agents. Furthermore, we extend MAPD to a customizable version, which can quantify differences among agent policies on specified aspects. 
Based on the online deployment of MAPD, we design a multi-agent dynamic parameter sharing (MADPS) algorithm as an example of the MAPD's applications. Extensive experiments demonstrate that our method is effective in measuring differences in agent policies and specific behavioral tendencies. Moreover, in comparison to other methods of parameter sharing, MADPS exhibits superior performance.

\end{abstract}


\newcommand{\BibTeX}{\rm B\kern-.05em{\sc i\kern-.025em b}\kern-.08em\TeX}

\makeatletter
\gdef\@copyrightpermission{
	\begin{minipage}{0.3\columnwidth}
		\href{https://creativecommons.org/licenses/by/4.0/}{\includegraphics[width=0.90\textwidth]{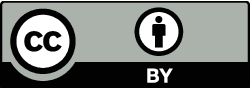}}
	\end{minipage}\hfill
	\begin{minipage}{0.7\columnwidth}
		\href{https://creativecommons.org/licenses/by/4.0/}{This work is licensed under a Creative Commons Attribution International 4.0 License.}
	\end{minipage}
	\vspace{5pt}
}
\makeatother

\begin{document}


\pagestyle{fancy}
\fancyhead{}


\maketitle 


\section{Introduction}
Multi-agent reinforcement learning (MARL) has achieved significant success in practical applications, such as multiplayer gaming~\cite{AlphaGO}, multi-robot system controlling~\cite{Multi_Robot}, sensor networks~\cite{Sensor_Networks} and autonomous driving~\cite{AutoCar}. Traditional MARL approaches commonly employ parameter sharing techniques to enhance sample efficiency~\cite{E,QMIX} and algorithm scalability~\cite{Transfer}. Although parameter sharing has been shown to be useful for accelerating training~\cite{SePS}, its excessive use can restrict the multi-agent system to fully shared policies, thus limiting their adaptability to complex tasks~\cite{CDS}. 

\begin{figure}[h]
  \centering
  \includegraphics[width=0.7\linewidth]{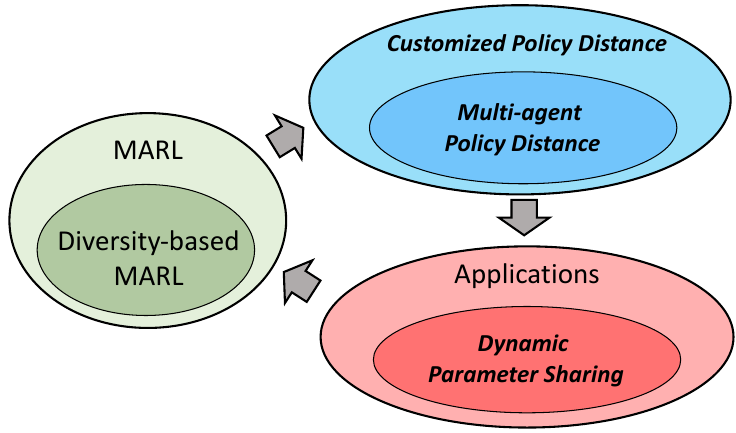}
  \caption{The relationship between our work and MARL. Our contributions are highlighted in bold and italicized.}
  \label{fig:our contributions}
  \Description{Action is not good.}
\end{figure}

The fundamental cause of this issue lies in the lack of policy diversity in multi-agent systems. This lack of diversity not only hinders the adaptability of multi-agent systems to complex tasks, but also weakens the overall exploration during learning. To address this problem, diversity-based MARL methods have emerged. One of the classic approaches to achieving policy diversity is to directly incorporate agent indices into the observations~\cite{F,G}. This allows the agents to exhibit different behaviors while sharing network parameters. Other intuitive approaches involve grouping the agents based on prior knowledge or learned information, and only sharing parameters within each group. For instance, ~\cite{HetGPPO} leverages the heterogeneous characteristics of agents for grouping, ~\cite{SePS} utilizes supervised learning and clustering to group the agents before training. Additionally, some methods introduce the concept of roles into MARL, controlling the parameters of the agents based on their roles. ~\cite{ROMA} captures emergent roles from agents' local observations, ~\cite{RODE} learns roles by embedding decomposed action spaces in the initial training phase, and ~\cite{ROGC} learns roles during training and updates the role clusters at fixed time steps.

Although the above methods allow agents to emerge policy diversity, there is currently a lack of a general metric that can quantify the differences in agent policies and be applicable in various types of multi-agent systems. It is meaningful to measure such differences between agent policies, as it can help us analyze the diversity evolution in multi-agent systems during learning, and gain a deeper understanding of how MARL can benefit from diversity. 

There have been previous attempts at measuring policy differences in MARL. ~\cite{I} introduces a "policy distance" by sampling states and recording the ratio of different actions selected by the population. ~\cite{J} employs KL divergence to measure the distances between action-based roles, ~\cite{K} uses the maximum mean discrepancy to compare distributions over trajectories under different policies. ~\cite{L} formulates the concept of behavioral diversity as the discrepancies of occupancy measures and utilizes f-divergence. Such methods based on divergence metrics fail to satisfy the triangle inequality. This results in increased challenges in analyzing multi-agent policies and complicates further applications. Some researches have delved into the representations of policies. In ~\cite{M}, episodes generated by agent policies are mapped to real-valued vectors, and the Euclidean distances between these vectors is utilized as the policy distances. \cite{N} maps agent policies to a distribution, and employs KL divergence and one-dimensional Wasserstein distance to handle discrete and continuous action spaces, respectively. These approaches, which map the overall policies to low-dimensional vectors or distributions, inevitably result in significant information loss. The state-of-the-art method for measuring behavioral diversity at a system level is presented in ~\cite{SND}. This method calculates the behavioral distance between pairs of agents, and employs the average of these distances as an indicator of system diversity. However, this approach requires assumptions about the type of agent policies and presumes that the observation and action spaces of agents are identical, limiting its generality.

We are desired to develop a general tool for measuring policy difference among agents. This tool not only allows us to gain a deeper understanding of the impact of diversity in MARL, but also provides guidance for the design of diversity-based algorithms. We model agent policies as action distributions conditioned on observations, and discuss the limitations of directly using action distributions for measuring. Our key insight, is to learn latent conditional representations of agents' decisions under each observation, and measure the accumulation of differences between these representations. Additionally, we extend MAPD to a customizable version, to quantify policy differences on specific aspects which the users are interest in. Based on the online deployment of MAPD, we design multi-agent dynamic parameter sharing (MADPS) as a paradigm for the utilization of MAPD. MADPS is an algorithm that can automatically adjust the parameter sharing scheme of agents during training, helping multi-agent systems find the appropriate balance between policy diversity and parameter sharing. The relationship between our work and MARL is illustrated in Figure~\ref{fig:our contributions}.

Our contributions can be summarized as follows:
\begin{enumerate}[label=(\arabic*)] 
  \item Proposing MAPD, a generalized measure of policy differences for multi-agent reinforcement learning.
  \item Developing a customizable version of MAPD for measuring policy differences on specific aspects.
  \item Designing MADPS, a multi-agent dynamic parameter sharing algorithm based on the online deployment of MAPD.
  \item Demonstrating the utilization of MAPD through cases in the multi-agent particle environment (MPE)~\cite{MADDPG}, and applying MADPS on both MPE and StarCraft II micromanagement environment~\cite{SMAC} to showcase its superior performance.
\end{enumerate}

\section{Background}
In this section, we introduce partially observable markov game (POMG)~\cite{HetGPPO}, and give a general mathematical definition of agent policies to establish the foundation of policy distance measurement.

A partially observable markov game can be represented by a 8-tuple:
$$
\langle \mathcal{N}, \mathcal{S} , \{\Omega^i\}_{i\in \mathcal{N}}, \{O^i\}_{i\in \mathcal{N}},\{A^i\}_{i\in \mathcal{N}}, \{R^i\}_{i\in \mathcal{N}}, \mathcal{T}, \gamma \rangle,
$$
where $\mathcal{N} = \{1,\cdots, n\}$ defines the set of all agents, $\mathcal{S}$ is the global state space. For an agent $i \in \mathcal{N}$, $\Omega^i$ denotes the observation space, $O^i$ denotes the corresponding observation function, $A^i$ is the action space and $R^i:\mathcal{S} \times\left\{A_i\right\}_{i \in \mathcal{N}} \times \mathcal{S} \mapsto \mathbb{R}$ is the reward function. $\mathcal{T}: \mathcal{S} \times \left\{A_i\right\}_{i \in \mathcal{N}} \mapsto \mathcal{S}$ represents the global state transition function, and $\gamma$ refers to the discount factor.

At each time $t$, an agent $i$ receives a partial observation $o_i^t \sim O^i(s_t)$ from the environment, and takes an action $a_i^t \in A^i$ according to its policy, where $s_t \in \mathcal{S}$ is the global state of the environment. Each agent in a POMG aims to learn an optimal policy $\pi_i^{*}$ that, given an observation at a specific time step, outputs the optimal action. The optimal policy can be learned by maximizing the total expected reward accumulated by the agent. 

In order to establish a general agent policy difference metric, the first thing we need to do is to give a general definition of agent policy. An agent's policy can always be represented as \textbf{a conditional probability distribution of actions under condition $o$}, that is, $\pi_i \coloneq \pi_i(a|o)$. This definition naturally holds when agents adopt stochastic policies and take actions based on policy-based MARL methods. In the case of value-based approaches, the action-values can be converted into a probability distribution across all actions, e.g., through softmax. Even when deterministic policies are utilized, actions can still be represented as a probability distribution, with a 100\% probability assigned to the chosen action. According to this definition, the policies of agents with discrete actions are discrete conditional distributions, while those corresponding to continuous actions are continuous distributions.


\section{Measuring Policy Distance between Agents}
\label{section3}
In this section, we present the desired input-output format and properties
of MAPD. Following that, we discuss the reasons and methods for learning a conditional representation for the agent decision-making. Finally, we show how to utilize this representation to compute a policy distance.

\subsection{Analysis}

\textbf{Our Objective.} This work aims to develop a general tool, termed multi-agent policy distance (MAPD), to measure the policy difference between any two agents in a multi-agent system,
$$
MAPD: \{(\pi_i, \pi_j)\}_{i,j \in \mathcal{N}} \mapsto \mathbb{R}_{\geq 0},
$$
where $\pi_i, \pi_j$ represent the policies of two agents in a multi-agent system. When the policies $\pi_i, \pi_j$ are given, MAPD can output a non-negative scale value, indicating the degree of difference between these two policies. And we hope this computed scale value to possess some distance-like properties.

\textbf{Properties of MAPD.} For ease of exposition, we use $\pi_i$ and $\theta_i$ to denote the policy and the network parameters of an agent $i$, respectively. The policy distance between agent $i$ and $j$ is represented by $d_{ij}$, and the properties of MAPD are as follows:

\begin{itemize}
    \item Symmetry: \hspace{1.2cm} $ d_{ij} = d_{ji}$.
    \item Non-negativity:  \hspace{0.6cm} $ d_{ij} \geq 0$.
    \item Identicals of indiscernibility (policies): $d_{ij}=0$  $\iff$ $\pi_{i}=\pi_{j}$.
    \item Identicals of indiscernibility (parameters): $d_{ij}=0$ $\Rightarrow$ $\theta_i=\theta_j$.
    \item Triangle inequality:  \hspace{0.1cm}  $d_{ij} \leq d_{ik}+d_{kj}$.
\end{itemize}

Among the above properties, \textit{symmetry} implies that the policy distance between two agents is not affected by their order. \textit{Non-negativity} specifies that the minimum policy distance is 0, indicating that two policies are identical, which corresponds to \textit{identicals of indiscernibility}. In order to emphasize the distinction between the policies and parameters of agents, we divide the concept of \textit{identicals of indiscernibility} into two aspects: one relating to the policies and the other to the parameters.  A policy distance of 0 implies that the policies of two agents are identical, and thus their parameters are also the same, the reverse is not true because the output of the agent's network may not equal to that of the policy. Therefore, directly comparing parameters of the neural network won't necessarily indicate policy similarity~\cite{28,39}.

The \textit{triangle inequality} is applicable in comparing the policies of multiple agents. We consider this property to be of great importance, since it determines that MAPD is not isolated between two agents, but also considers the influence of other agents. This property allows us not only to estimate other policy distances based on a given set of distances, but also to guide the design of dynamic parameter sharing in Section~\ref{section:5.1}.

According to Prop. 2 and Prop. 3, a policy distance equals to 0 when the agent policies are identical, whereas a policy distance bigger than 0 when there are differences in policies. Thus, we need to think about what would cause an increase in the differences of policies, or in other words, an increase in policy distance between two agents. A straightforward approach is to compare the action distributions under the same observations, and accumulate the differences between these distributions. To capture even the slightest differences related to the overall policy, these calculations need to be conducted across the entire observation space. Therefore, a vanilla method~\cite{SND} of measuring policy distance can be obtained as follows:
\begin{eqnarray}
d_{ij}^{Vanilla}=\int_{\mathcal{O}} f\left[\pi_{i}(a|o), \pi_{j}(a|o)\right] d o,
\end{eqnarray}
where $\pi_{i}(a|o)$ denotes the action distribution of an agent $i$, $f$ is a certain metric for measuring the distance between two distributions, $\mathcal{O}$ is the observation space. 

However, \textbf{directly measuring the action distributions may not be an ideal approach} due to certain requirements related to the type of distribution. Experiments in ~\cite{SND} assume Gaussian-distributed actions in all continuous settings. Contrarily, agents might take actions under different distributions, such as a bimodal one. When the action distributions between two agents exhibit diverse forms or correspond to different action spaces, measuring the distance between these distributions becomes unfeasible.

Things may get worse in scenarios where agents take discrete actions. Consider a scenario with five discrete actions: given the same observation, an agent \textit{A} outputs an action distribution of $[\textbf{1.0},0,0,0,0]$, indicating a 100\% preference for action No.1. Agent \textit{B} and \textit{C} have distributions $[0,0,0,\textbf{1.0},0]$ and $[0,0,0,0,\textbf{1.0}]$, respectively. According to the definition of Wasserstein distance (WD)~\cite{Wasserstein} and Hellinger distance (HD)~\cite{Hellinger}, the WD between $\textit{A}$-$\textit{B}$ is three times as big as that between $\textit{B}$-$\textit{C}$ (3 and 1), while both $\textit{A}$-$\textit{B}$ and $\textit{B}$-$\textit{C}$ have identical HD of 1. The crux of this matter lies in the non-sequential nature of action indices, where index-close actions can denote either closely related or vastly different operations.

Hence, our idea is to identify a distribution $p_i(z|o)$ in a latent space to represent the agent's action distribution $\pi_{i} (a|o)$. Within this designated latent space, the inherent non-sequential nature can be eradicated, and the types of distributions can also be standardized. This can be achieved by assuming the latent variables following to a certain multi-dimensional prior distribution. We can measure distances between these distributions as an alternative to measuring action distributions, while the semantics of each dimension within this distribution is not of concern to us.

\begin{figure*}[h]
  \centering
  \includegraphics[width=0.5\linewidth]{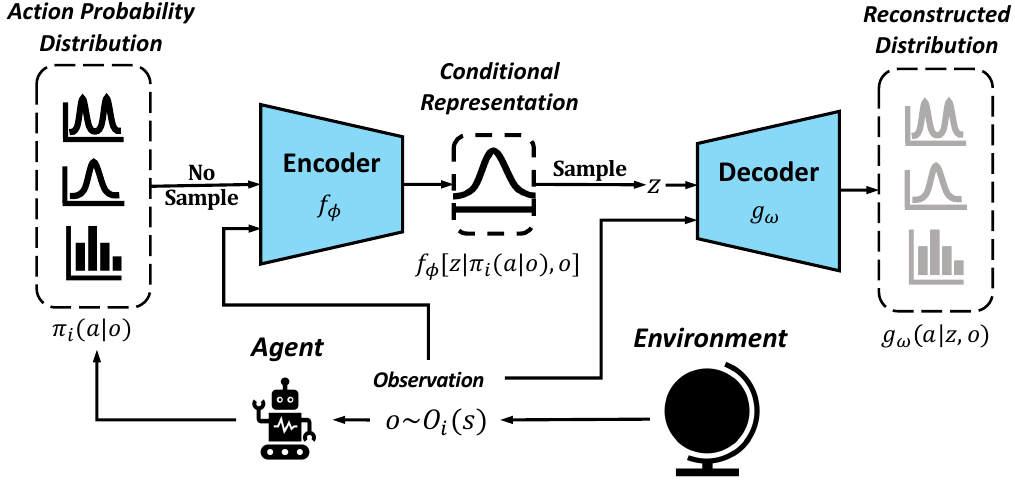}
  \caption{Learning the conditional representation of an agent's decision.}
  \label{fig:Auto-Encoder}
  \Description{Ada-HetGAT.}
\end{figure*}

\subsection{Learning the conditional representations of agents' decisions}
\label{section-3.2}

We hope to find a mapping function, which can output a latent distribution when the observation $o$ and an agent $i$'s decision $a_i \sim \pi_{i} (a|o)$ are given. Thus, this latent distribution can be presented as a posterior $q(z|a_i,o)$ under $o$ and $a_i$, where $q$ denotes the desired mapping function.

In order to ensure the representational capability of $q(z|a_i,o)$, we employ an auto-encoder architecture to learn $q$. The encoder, which approximates the posterior distribution, takes $a_i$ and $o$ as input and outputs a distribution of the latent variable $z$. On the other hand, the decoder tries to reconstruct $a_i$ by drawing samples $z$ from $q(z|a_i,o)$. Solely using $z$ as input is insufficient since the agent's actions inherently depend on $o$, so the decoder also incorporates \(o\) to reconstruct the action.

However, the auto-encoder framework is widely used as a generative model~\cite{Diffusion}, which predicts the underlying distribution of sample data (in our case $a \sim \pi_{i} (a|o)$). Since the input action $a$ is sampled from the agent's action distribution, it requires an ample amount of data to fit the action distribution $\pi_{i} (a|o_x)$ for agent \(i\) under all possible observations \(o_x\), where \(x\) denotes the index of observation. 

Fortunately, the agent's policy is fixed during the training of the auto-encoder, meaning the action distributions for each observation is known. Thus, we directly use $\pi_{i} (a|o)$ as the encoder's input and explicitly incorporate agent's observation \(o\). The decoder no longer reconstructs the action codes but the action distribution. In this case, the conditional latent variable distribution changes from \(q(z|a_i,o)\) to a posterior \(q[z|\pi_{i}(a|o),o]\).

The architecture for learning this conditional representation is illustrated in Figure~\ref{fig:Auto-Encoder}. An agent $i$ receives observation $o\sim O_i(s)$ through its observation function and outputs the corresponding action distribution \(\pi_{i}(a|o)\), where $O_i(s)$ represents the observation function, $s$ is the global state. The encoder $f$, parameterized by $\phi$, accepts inputs $\pi_{i}(a|o)$ and $o$, and produces the parameters of the latent variable's posterior probability distribution. The decoder $g$ uses the sampled \(z\) and \(o\) to yield the reconstructed action distribution, with its parameters denoted by \(\omega\). This auto-encoder model is trained by minimizing the following loss function:
\begin{equation}
\begin{aligned}
\mathcal{L}\left(\phi, \omega \right)
&= \mathbb{E}_{z \sim f_\phi [z|\pi_{i}(a|o),o]} \left[ D_{\Pi} \left( g_{\omega} (a|z,o)  || \pi_{i} (a|o)  \right) \right] \\
&\quad + D_{K L} \left( f_\phi [z|\pi_{i}(a|o),o] || p(z|o)\right),
\end{aligned}
\label{MAPD-loss}
\end{equation}
where the first term is the reconstruction loss, and the second term is the KL loss for bring generated latent distribution $f_\phi [z|\pi_{i}(a|o),o]$ closer to the prior \(p(z|o)\). In our experiments, \(p(z|o)\) is a multi-dimensional Gaussian distribution.

In the reconstruction loss term, $D_{\Pi}(x||y)$ is a generalized distribution distance which measures the similarity between two distributions $x$ and $y$. In Equation~\ref{MAPD-loss}, $x$ refers to the reconstructed action distribution $g_{\omega} (a|z,o)$ and $y$ is the original action distribution $\pi_{i}(a|o)$. The specific form of $D_{\Pi}$ depends on the form of $\pi_{i}(a|o)$. A question may arise: since learning this representation is to avoid directly measuring the distance between agents' action distributions, why the (action distribution) distance reoccurs here? In fact, this measurement only ensures a close alignment of the reconstructed action distribution with the original one, without requiring the various properties necessary for MAPD. Therefore, $D_{\Pi}$ can be a WD (standard distance metrics), a KL divergence (f-divergences), or even the MSE Loss of the action distribution vectors.

\begin{figure*}[h]
  \centering
  \includegraphics[width=0.8\linewidth]{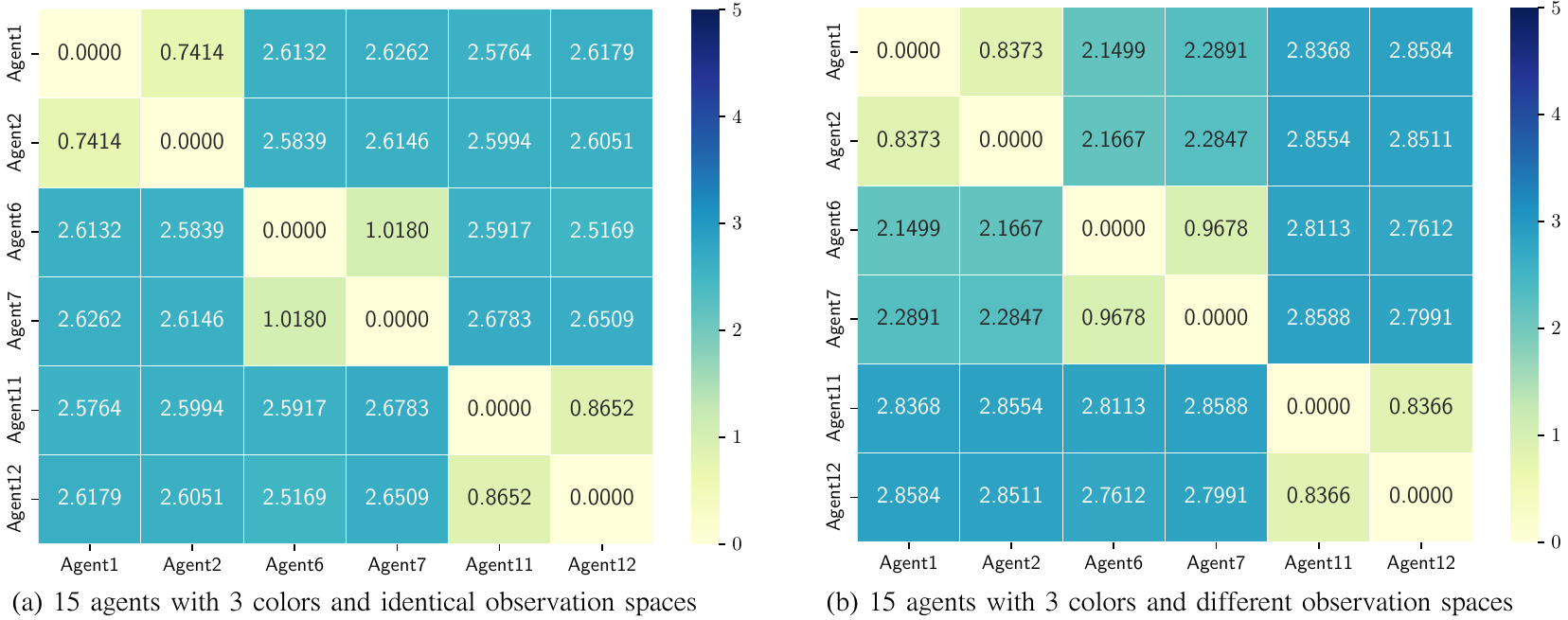}
  \caption{Policy distance matrices in multi-agent spread tasks. In this scenario, there are 15 agents with 3 colors: agents numbered 1-5, 6-10, and 11-15 are each given the colors No.1, No.2, and No.3. The agents must move to the specific landmarks that matches their color. Theses matrices show the policy distances just for the first two agents in each of colored groups.}
  \label{fig:MAPD-case}
  \Description{policy_distance_example}
\end{figure*}

\subsection{Multi-Agent Policy Distance}

After obtaining the conditional representations of an agent's decisions, MAPD uses these learned distributions to replace the action distributions under each observation. Since the action distributions are mapped into a latent distribution in a same feature space, we are able to measure the policy distance between a pair of agents which have different types of actions. Considering that the observation spaces may also be different, MAPD measures the policy differences between two agents on a common space $\mathcal{O}_P$, which is a set formed by padding the observations of these two agents. The policy distance between two agents following MAPD can be computed as follows:
\begin{eqnarray}
d_{i j}=\int_{\mathcal{O'} \subseteq \mathcal{O}_P} W\left[p_i(z|o), p_j(z|o)\right] d o,
\label{equation:MAPD}
\end{eqnarray}
where $p_i(z|o) = f_\phi [z|\pi_{i}(a|o),o]$, denotes the latent distribution of agent $i$ under observation $o$, $W$ denotes a distribution distance, $\mathcal{O'}$ is a subset of the observation space $\mathcal{O}_P$. More details about Equation~\ref{equation:MAPD} can be found in Appendix 1.1.


In practical applications, it is intractable to compute the integral by traversing the observation space. In fact, the set of observations involved in multi-agent tasks (determined by the initial state and the state transition function) may only constitute a small portion of the entire observation space. Hence, we focus only on the relevant observations $o\in\mathcal{O'}$, specifically those that occur in the actual tasks. In experiments, we use Monte Carlo sampling to estimate the integral and employ multi-rollout to reduce estimation variance.

\subsection{Case Study of MAPD}
\label{section-3.4}
In this section, we use a multi-agent spread task~\cite{MADDPG} as a case study to demonstrate the effectiveness of MAPD. In this scenario, there are 15 agents with 3 colors: the first 5 agents are assigned color No.1, the next 5 and last 5 are assigned color No.2 and No.3, respectively. Each agent must move exclusively towards the landmark matching its assigned color. The positions of the agents and landmarks are randomly initialized, and the agents need to avoid collisions with other agents, while moving towards the corresponding landmarks. The neural networks of all agents are set to be independent, and the parameters are initialized as the same.

Figure~\ref{fig:MAPD-case} shows parts of the policy distance matrices in two scenarios. In scenario (a), the observation spaces of agents are the same, while in scenario (b), the observations are randomly shuffled, resulting in different observation spaces. Both matrices are obtained at 10000 steps during training over 4 random seeds, with only 100 steps of data used for auto-encoder training and policy distance computation. We can observe that each agent has a policy distance of 0 with itself, corresponding to Prop. 3 of MAPD. Agents of the same color should move to the same landmark (similar policies), as indicated by their policy distances being less than 1.1. 

On the other hand, agents of different colors, due to their different tasks,  need to move to different landmarks even under the same observation, resulting in their policy distances all bigger than 2.1. Furthermore, this phenomenon exists regardless of whether the agents have the same observation space. This indicates that the proposed MAPD is able to effectively capture and quantify the differences in agent policies.


\section{Measuring Customized Policy Distance}
\label{headings}

In some cases, beyond identifying the overall difference between two agent policies, we may also be curious about their differences in certain aspects. Agents might have distinct patterns in some aspects of policy, while exhibiting similar patterns in others. Measuring the \textit{policy distance in certain aspects} can help us determine in which areas the agents truly differ, and in which they show similar inclinations.

We refer to this special policy distance as the \textbf{customized policy distance}. In this section, we will introduce our method of measuring customized policy distance, and give a case study to illustrate its application.

\subsection{Learning customized representations}

In fact, it is challenging to define and quantify certain aspects of policies.
This is because this abstract concept is manually defined, rather than just being a part of the action or the policy. However, these manually defined aspects can be linked to feedback from environments or agents in MARL. For example, if we want to measure the differences in \textit{aggressive} and \textit{defensive} aspects of units' policies in StarcraftII~\cite{SMAC}, we can find the underlying \textit{bonds} between their actions and the "killing reward" and "defensing reward" in the environment, and then use these bonds to measure the differences.

We refer to such information from environments or agents as \textbf{customized features}, and the bond as \textbf{customized representations}. Customized features possess the following properties:

\begin{itemize}
    \item Customized features can be modeled as stochastic variables.
    \item Customized features follow a conditional probability distribution based on observations.
\end{itemize}

The process of measuring the customized policy distance is similar to the methodology of measuring policy distance in Section~\ref{section3}. Both need to learn a conditional representation of agent's decisions, but the customized one no longer reconstruct the policy itself, but to predict the conditional distribution of the customized features. For a clearer exposition of customized feature and customized policy distance, we denote the customized feature as $\boldsymbol{c}$, which follows some probability distribution $p(\boldsymbol{c}|o)$ conditioned on observation. In order to learn a meaningful latents representing the customized feature, our objective is to maximize the likelihood $\log p(\boldsymbol{c}|o)$ of all observed $\boldsymbol{c}$ under condition $o$~\cite{Diffusion}.

To optimize the $\log p(\boldsymbol{c}|o)$, an Evidence Lower Bound (ELBO) of the likelihood can be derived as follows:
\begin{equation}
\begin{aligned}
\log p(\boldsymbol{c}|o) 
& \geq \mathbb{E}_{f_{\phi^c}[z|\pi_{i}(a|o),o]} 
\left[ \log
\frac{p(\boldsymbol{c},z|o)}{f_{\phi^c}[z|\pi_{i}(a|o),o]}
\right],   \\
\end{aligned}
\label{ELBO}
\end{equation}
where $f_{\phi^c}[z|\pi_{i}(a|o),o]$ represents the posterior probability distribution of the latent variable generated by the encoder, and $p(\boldsymbol{c},z|o)$ denotes a joint probability distribution concerning the customized feature and latent variable, conditioned on $o$. 

Considering that the ELBO includes an unknown joint probability distribution, we can further decompose it by using the posterior probability distributions from the encoder and decoder:
\begin{equation}
\begin{aligned}
& \mathbb{E}_{f_{\phi^c}[z|\pi_{i}(a|o),o]} 
\left[ \log
\frac{p(\boldsymbol{c},z|o)}{f_{\phi^c}[z|\pi_{i}(a|o),o]}
\right] \\
& =\underbrace{\mathbb{E}_{f_{\phi^c}[z|\pi_{i}(a|o),o]} \left[\log
g_\omega(\boldsymbol{c}|z,o)\right]}_{\text {reconstruction term }} 
-
\underbrace{D_{\mathrm{KL}}\left(f_{\phi^c}[z|\pi_{i}(a|o),o]
 \| p(z|o)\right)}_{\text {prior matching term }} .\quad
\\
\end{aligned}
\label{ELBO-decompose}
\end{equation}

In Equation~\ref{ELBO-decompose}, $f_{\phi^c}[z|\pi_{i}(a|o),o]$ and $g_\omega(\boldsymbol{c}|z,o)$ are the posteriors from the encoder and decoder, respectively. The detailed derivation can be found in Appendix 1.2. Thus, the ELBO can be decomposed into a reconstruction term of the customized feature, and a prior matching term of the posterior and the prior. For the training of the auto-encoder, we utilize both the reconstruction and matching components as loss functions.

It is not difficult to notice that the form of Equation~\ref{ELBO-decompose} is very similar to that of Equation~\ref{MAPD-loss}. As a matter of fact, the normal policy distance is a special case of the customized distance when $\boldsymbol{c}$ is the action, as the actions follow a conditional distribution based on observations. Furthermore, the theory of customized distance demonstrates the rationality of the loss design in Section~\ref{section-3.2}. Similar to MAPD, the customized distance between agents can be computed as follows:

\begin{eqnarray}
d_{i j}^c=\int_{\mathcal{O'} \subseteq \mathcal{O}_P} W\left[p_i^c(z|o), p_j^c(z|o)\right] d o,
\end{eqnarray}
where $d_{i j}^c$ denotes the customized policy distance between agent $i$ and agent $j$, $p_i^c(z|o) = f_{\phi^c}[z|\pi_{i}(a|o),o]$ is the customized representations of agent $i$’s decisions.

\subsection{Case Study of Customized MAPD}
\label{section:4.2}
To give an example of measuring the customized policy distance, we refer to the multi-agent spread scenarios in Section~\ref{section-3.4}. In this section, we measure the policy distances of the multi-agent system based on two aspects: 1) \textit{The tendency of moving towards a same landmark} and 2) \textit{The tendency of moving towards the matching landmark}.

In this case, we choose the distance $dis_i$ between an agent $i$ and the landmark as the customized feature, to link agent's decisions and the specific aspects. This distance is sampled at step $t+1$ corresponding to the observation $o_i^t$. Thus, the likelihood term in Equation~\ref{ELBO} corresponds to $p(\boldsymbol{c}|o) = p(dis_i^{t+1}|o_i^t)$.

\begin{figure*}[h]
  \centering
  \includegraphics[width=0.8\linewidth]{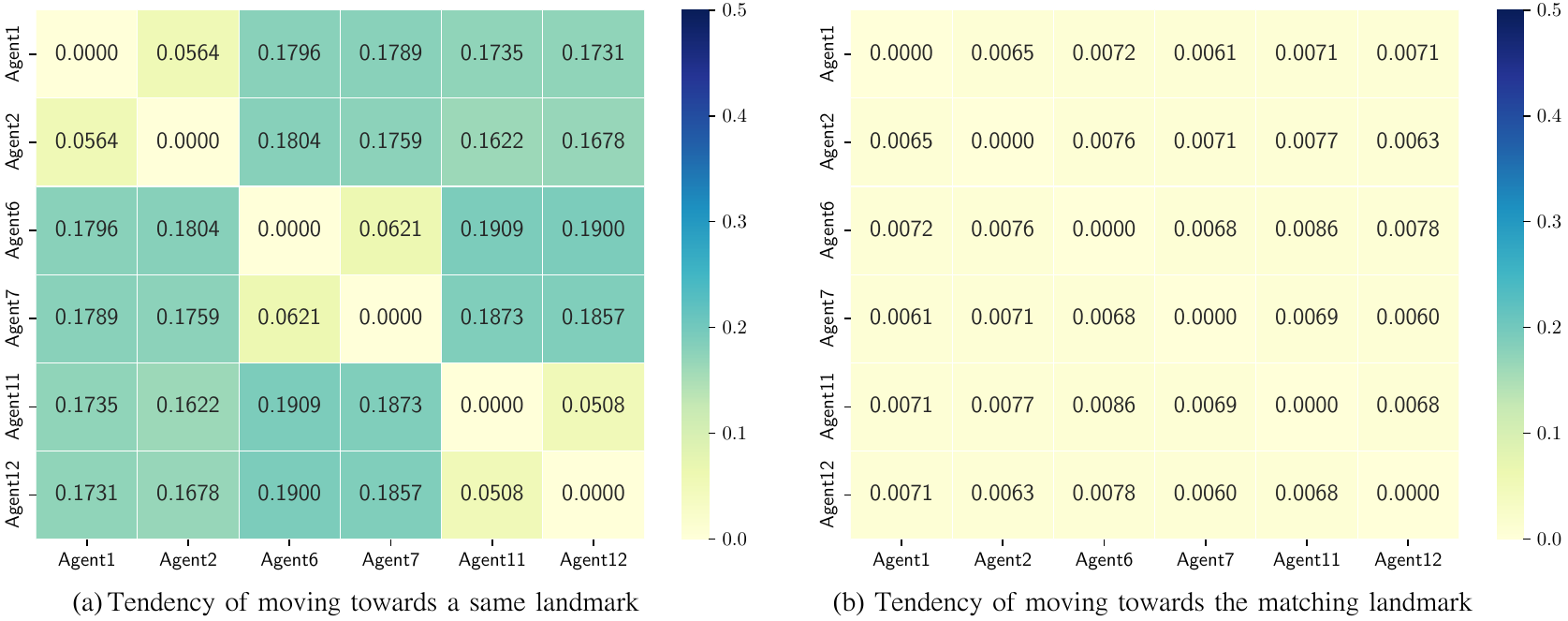}
  \caption{Customized policy distance matrices in multi-agent spread tasks. Figure (a) demonstrates the policy distances between agents on the tendency of \textit{moving towards a same colored landmark}, figure (b) demonstrates the policy distances on the tendency of \textit{moving towards the matching landmark}.}
  \label{fig:CMAPD-case}
  \Description{customized_distance_example}
\end{figure*}

As for the experiments, we use the same parameter settings in Section~\ref{section-3.4}, and record customized policy distances of the multi-agent system trained for 10000 steps (only 100 steps of data for auto-encoder training and policy distance computation), the results are shown in Figure~\ref{fig:CMAPD-case}.

It can be observed that, when considering the tendency of moving towards a same landmark (in our experiments the landmark No.1), the customized policy distances varies significantly depending on the colors of agents. The customized policy distances among agents with different colors are three times as big as that among agents with the same colors. This ratio is similar to the ratio of policy distances shown in Figure~\ref{fig:MAPD-case}, indicating that the tendency of spreading towards landmark No.1 is one of the factors contributing to the differences in agent policies.

As for the tendency of moving towards the matching landmarks, all the customized policy distances become extremely small (all less than 0.01). This result is reasonable because all agents have the same goal of moving towards landmarks of their own color, thus they are consistent in this aspect. These results demonstrate the effectiveness of our proposed method, in quantifying the differences in certain aspects of agent policies.

\section{Dynamic Parameter Sharing: an application of MAPD for MARL}
\label{others}

Thanks to the low data requirement of MAPD, it becomes possible to deploy MAPD online during the training of multi-agent systems. Having a precise online measurement will enable a real-time evaluation of the policy diversity evolution within the multi-agent system, and such evolution can be intervened if needed. This section provides an application of online deployment of MAPD, multi-agent dynamic parameter sharing (MADPS).

\subsection{Multi Agent Dynamic Parameter Sharing}
\label{section:5.1}
The motivation of MADPS comes from the need for a balance between policy diversity and parameter sharing in multi-agent systems. As parameter sharing can improve the efficiency of MARL, it can also constrain the policy diversity of multi-agent systems, and consequently affect the performance of MARL. If a dynamic parameter sharing can be accomplished during training and agents' policies are appropriately shared when necessary, the multi-agent system can not only adapt to tasks with emergent diversity, but also benefit from the sample efficiency of parameter sharing.

\begin{figure}[h]
  \centering
  \includegraphics[width=0.99\linewidth]{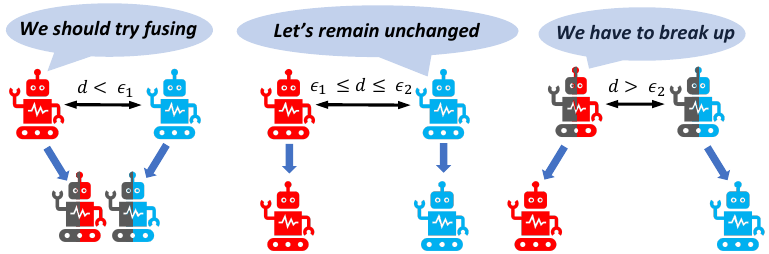}
  \caption{The basic idea of dynamic parameter sharing.}
  \label{fig:basic idea of MADPS}
  \Description{Action is not good.}
\end{figure}

The basic idea of MADPS is to \textit{fuse} the policies of closely related agents while \textit{dividing} the policies of agents with significant differences. We employ the normal/customized policy distance to measure the similarity between agent policies, and conduct a dynamic \textit{fusion} and \textit{division} of agent policies. During each fusion operation, only a portion of the network parameters is shared, to prevent multi-agent policies from getting stuck in local optima. Similarly, during each division operation, we only split a portion of the network parameters. MADPS uses a fusion threshold $\epsilon_1$ and a division threshold $\epsilon_2$ to determine whether fusion or separation should be conducted, as shown in Figure~\ref{fig:basic idea of MADPS}.

However, when addressing parameter sharing among multiple agents, the interrelationship between $\epsilon_1$ and $\epsilon_2$ and the order of fusion and division should be considered carefully. Based on the triangular inequality property of MAPD, we indicate that the division threshold should satisfy $\epsilon_2 \geq 2\epsilon_1$. In the experiments, we directly set $\epsilon_2=2\epsilon_1$ to reduce the number of hyperparameters. Additionally, an algorithm for the fusion and division of multi-agent policies is developed, in order to avoid various corresponding contradictions (refer to Appendix 2 for algorithm details and relevant explanations).

Thus, we introduce this algorithm of multi-agent policy fusion and division in MARL, resulting in MADPS. MADPS allows a multi-agent system to begin training with flexible forms of parameter sharing initialization (e.g., fully shared policies, independent policies, or selective parameter sharing~\cite{SePS}.) In every $T$ time steps, MADPS computes a policy distance matrix of the multi-agent system. By comparing the policy distances with predefined thresholds, MADPS conducts appropriate fusion and division of policies, until a stable parameter sharing structure is achieved, striking a balance between parameter sharing and policy diversity.

\begin{figure*}[h]
  \centering
  \includegraphics[width=0.85\linewidth]{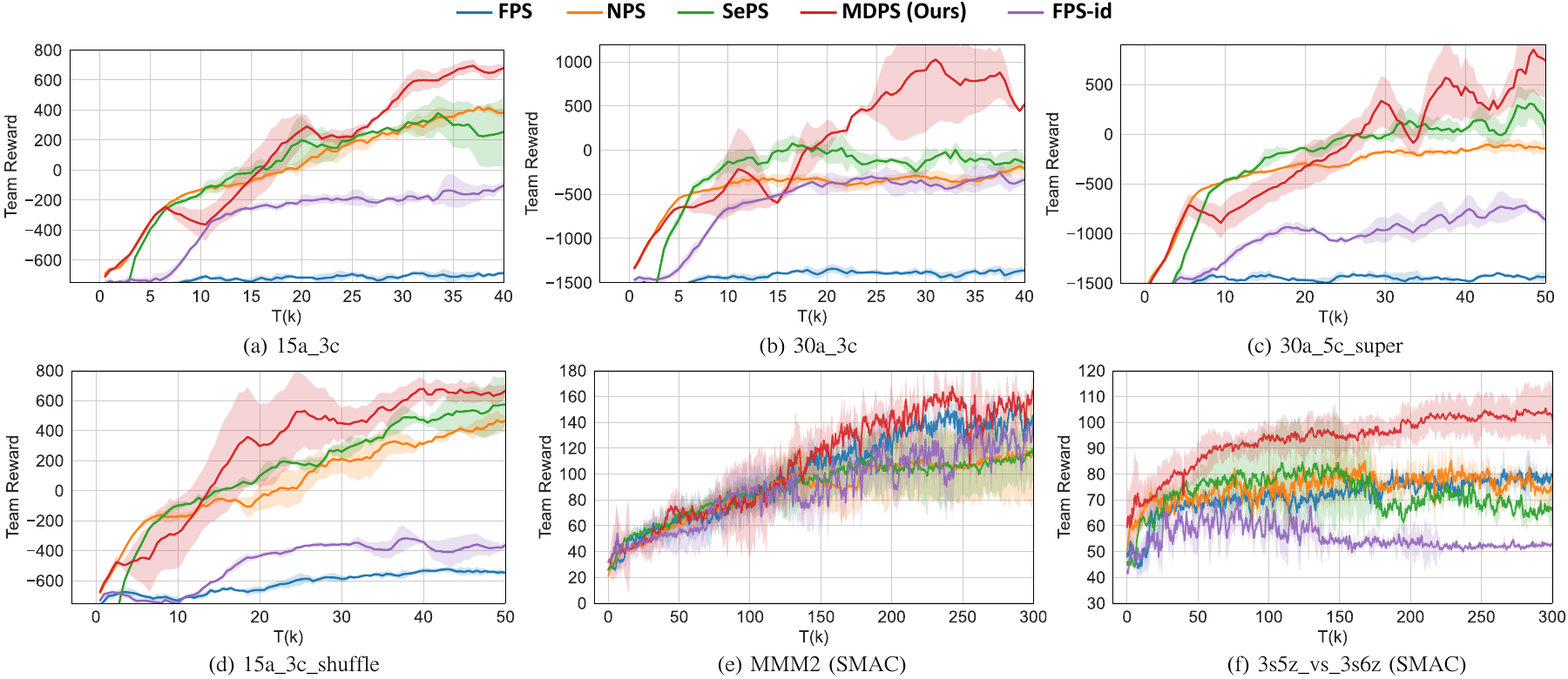}
  \caption{Performance comparison with baselines on a series of multi-agent spread tasks and super-hard tasks in SMAC.}
  \label{fig:exp-results}
  \Description{Ada-HetGAT.}
\end{figure*}

\section{Experiments}
In Section~\ref{section-3.4} and Section~\ref{section:4.2}, we employ a toy environment to demonstrate the effectiveness of MAPD. In this section, the online deployment of MAPD and MADPS are applied to more challenging multi-agent spread tasks and StarCraft II micromanagement environment (SMAC). By comparing MADPS with other methods of parameter sharing, we analyze the performance improvement of MARL algorithms through appropriate policy diversity and parameter sharing, and demonstrate the superior performance of MADPS based on MAPD. The code is available here\footnote{\href{https://github.com/Harry67Hu/MADPS}{https://github.com/Harry67Hu/MADPS}}.

\subsection{Experiment Settings}
\textbf{Multi-agent environments}. This part introduces the testing environment for our method. We increase the difficulty of the original multi-agent spread task, by requiring the agents not only to spread towards corresponding landmarks, but also to gather as closely as possible once they reach the landmarks. Besides the original penalties for collisions and guidance, a sparse reward for encourage agents moving closely is added in the overall rewards. This sparse reward is only given to agents when the distance between the agents and the landmark is smaller than a threshold, increases as the distance between the agents decreases, and decreases with the number of agents (which meeting the threshold condition) decreases.
Another testing environment used in this study is SMAC. All the testing methods are evaluated on four super-hard maps in SMAC, namely, \textit{corridor}, \textit{6h\_vs\_8z}, \textit{3s5z\_vs\_3s6z} and \textit{MMM2}. The first two maps involve homogeneous agent scenarios, while the last two represent scenarios with heterogeneous agents.

\textbf{Baselines}.
We compare the proposed dynamic parameter sharing with other methods of parameter sharing. To ensure fairness, all methods employ the same actor-critic architecture for training, and the network structures and hyperparameter settings are kept identical. (More details can be found in Appendix 3.2). The baselines can be summarized as follows:

\begin{itemize}
    \item No parameter sharing (NPS): each agent has an independent network with no parameter sharing.
    \item Fully parameter sharing (FPS): all agents fully share the same network parameters.
    \item Fully parameter sharing with index (FPS-id): all agents share the network parameters, but a unique index is attached to each agent's observation.
    \item Selective parameter sharing (SePS): before the training of MARL, agents are clustered according to learned latent representations, and parameter sharing is restricted to each agent cluster~\cite{SePS}.
\end{itemize}

\subsection{Superior Performance of Our Method}
Our method and the baselines are tested on four super-hard maps in SMAC and a series of harder multi-agent spread tasks. The variations in multi-agent spread tasks include the number of agents, the number of agent colors, the allocation of agent colors, and the structure of agent observations. 

We represent a scene with \( x \) agents and \( y \) colors as \textit{xa\_yc} , and use \textit{super} to denote scenes where agent colors are not evenly allocated, and \textit{shuffle} to indicate scenes where agent observations are randomly shuffled. For example, \textit{15a\_3c\_shuffle} represents a scene with 15 agents, where the agents are evenly allocated among 3 colors based on their indices (the agents numbered 1-5, 6-10, and 11-15 are given the colors No.1, No.2, and No.3, respectively), and the agent observations are shuffled.

Figure~\ref{fig:exp-results} illustrates the performance of all methods on a selection of the above environments. More experimental results can be found in Appendix 3.2. 

\balance

It can be observed that FPS cannot learn effective policies in all multi-agent spread tasks. Because in these scenarios, agents of different colors move in different directions due to guided penalties, but these experiences are inappropriately used in the training of the shared network, causing the policy to fluctuate between different goals and leading to training failure. The curves of FPS-id can converge to a value close to 0 in scenarios (a) and (b), but they still fail to trigger the designed sparse rewards. Additionally, in the \textit{shuffle} scenario (d), the shared network cannot effectively recognize the types of agents due to the shuffled observations, thus preventing the learning of effective policies.

On the other hand, NPS can achieve a mediocre performance on all tasks. In multi-agent spread tasks, this approach actually allows agents to learn unique policies, thus achieving a overall behavior diversity in multi-agent systems. But in other tasks such as \textit{MMM2}, 
the training of NPS is slow due to its independent network setting, and it cannot use the experience of other agents to train networks.

By analyzing the curves of NPS, FPS, and FPS-id on different tasks, we can gain a rough understanding of how important policy diversity is in each task. For instance, in all multi-agent spread tasks, policy diversity is crucial for improving algorithm performance, but excessive policy diversity can hinder agents from achieving higher rewards. Additionally, the performance difference between FPS and FPS-id is much smaller in \textit{MMM2} compared to \textit{3s5z\_vs\_3s6z}. This indicates that although the both are heterogeneous scenarios, the task of \textit{3s5z\_vs\_3s6z} requires more policy diversity for agents than \textit{MMM2}.

Both MADPS and SePS encourage policy diversity among agents, but our MADPS uses a smarter approach. SePS utilizes each agent's trajectory to learn a latent representation of its index, and employs this representation for clustering, thus dividing agent network parameters based on groups. However, such grouping is fixed during training, and may be overly deliberate, resulting in less favorable experimental performance in some scenarios. To ensure a good performance, SePS requires prior knowledge of each scenario, necessitating a setting of unique hyperparameters for each scenario. On the other hand, MADPS only employs one hyperparameter across a series of scenarios, and can automatically adjust the parameter sharing scheme during training.

These experimental results demonstrate the effectiveness of the idea, i.e., dynamically adjusting parameter sharing in MARL. The essence behind this idea is the indispensable need for appropriate policy diversity as well as parameter sharing techniques in MARL. Thanks to our proposed MAPD, we are able to accurately capture the differences in policies between agents. This enabling a further understand of policy diversity for specific tasks, and provides valuable guidance for the design of MARL algorithms in practical applications.


\section{Conclusion}

This paper proposes MAPD, a method of quantifying differences in policies of MARL. Instead of directly measuring the polices, MAPD learns conditional representations of agents' decisions, and computes the policy distance by integrating the distances of the latent distributions. This approach is applicable to various policy forms of multi-agent systems. The variant of MAPD, customized policy distance, allows for measuring the behavioral differences between agents on specific aspects of policy. Based on MAPD, we design a dynamic parameter sharing algorithm, which can achieve a dynamic fusion and division of agents' policies during the MARL training. Experimental results demonstrate that our method can accurately quantifiy the diversity in agent policies, and MADPS outperforms other baseline methods of parameter sharing. We believe that MAPD can serve as an important tool for studying and leveraging behavioral diversity in multi-agent systems, ultimately promoting the advancement of MARL.


\balance




\begin{acks}
This work was supported by the Strategic Priority Research Program of Chinese Academy of Sciences under Grant XDA27030204, the National Natural Science Foundation of China under Grant 62073323, and the Beijing Nova Program under Grant 20220484077.
\end{acks}



\bibliographystyle{ACM-Reference-Format} 
\bibliography{Measuring_Policy_Distance_for_MARL/Measure}

\end{document}